 \documentclass[final,5p,times,twocolumn,authoryear]{elsarticle}

\usepackage{amssymb}
\usepackage{lipsum}
\usepackage{amsmath}

\journal{High Energy Astrophysics}

\begin{document}

\begin{frontmatter}



\title{Reconnection-Driven Injection and Stochastic Reacceleration during Cosmological Magnetogenesis}


\author[first]{Ji-Hoon Ha}
\affiliation[first]{organization={Korea Astronomy and Space Science Institute},
            addressline={776 Daedeok-daero, Yuseong-gu}, 
            city={Daejeon},
            postcode={34055}, 
            country={Republic of Korea}}

\begin{abstract}
We investigate whether magnetic reconnection can provide suprathermal proton seed particles during cosmological magnetogenesis prior to nonlinear structure formation. Previous work showed that pressure-anisotropy-driven stochastic acceleration alone is strongly limited by cosmological expansion and Coulomb cooling. Here, we extend this framework by adding a phenomenological reconnection-driven source term to the Fokker--Planck equation for the isotropic ion distribution, with the injection power tied to the magnetic-energy growth rate during magnetogenesis. We find that reconnection can act as a fast injection channel and can produce a visible suprathermal tail. However, the resulting nonthermal energy fraction remains very small, of order $10^{-7}$ in the fiducial model, implying a negligible nonthermal pressure contribution to the pre-structure intergalactic medium. This limitation arises because the extremely high-beta plasma contains only a small magnetic-energy reservoir, even when reconnection itself is locally fast. Using a test-particle shock reacceleration estimate, we further show that the reconnection-produced tail can enhance the suprathermal proton population available for later structure-formation shocks by about an order of magnitude. Nevertheless, the associated hadronic gamma-ray emission from low-density cluster outskirts is expected to remain far below current detectability. We therefore conclude that reconnection during cosmological magnetogenesis is unlikely to dominate the cosmic-ray energy budget directly, but may provide a low-level seed population for subsequent shock acceleration.
\end{abstract}




\end{frontmatter}




\section{Introduction}

Cosmic rays (CRs) are an important nonthermal component of astrophysical
plasmas, and their production is usually associated with collisionless shocks.
In large-scale structure formation, shocks generated by hierarchical collapse
are expected to accelerate particles through diffusive shock acceleration and
to contribute to the CR content of the intracluster and intergalactic media
\citep[e.g.,][]{Miniati2000,Ryu2003,Pfrommer2006,Hoeft2008,Skillman2008,
Vazza2009,Hong2014,Schaal2015,Ha2018a,Ha2020,Ha2023}. However, the origin of
the first suprathermal seed particles before the widespread emergence of
structure-formation shocks remains less clear. If a weak pre-existing
suprathermal population is already present, it may affect the subsequent
injection efficiency at later shocks, even if it does not dominate the total
CR energy budget.

A possible pre-shock acceleration channel is associated with cosmological
magnetogenesis. In weakly magnetized plasmas, pressure-anisotropy-driven
microinstabilities can amplify magnetic fields and enhance pitch-angle
scattering \citep[e.g.,][]{Schekochihin2005,Schekochihin2006a,
Schekochihin2006b,Falceta2015,Ha2025}. As the magnetic field grows, the ion
gyrofrequency increases, potentially reducing the acceleration time of
stochastic, second-order Fermi processes. This possibility was investigated in
our previous work, where we considered stochastic acceleration during
pressure-anisotropy-driven magnetogenesis and solved the corresponding
Fokker--Planck equation including cosmological expansion and Coulomb losses
\citep{Ha2026}. The main result was that stochastic acceleration alone is
strongly limited in the pre-structure epoch. Although instability-enhanced
scattering can provide some pre-acceleration, Coulomb cooling and the Hubble
expansion suppress the formation of a dynamically significant nonthermal
population.

This conclusion leaves an important possibility open. 
The inefficiency of the stochastic channel does not exclude faster injection
mechanisms that can first inject particles from the thermal pool into the
suprathermal regime. Magnetic reconnection within turbulent current sheets is
a natural candidate for such an injection process. 
The relevance of magnetic reconnection in large-scale cosmic plasmas, including turbulent intracluster plasmas, has long been discussed in connection with plasma heating, particle acceleration, and cluster-scale nonthermal activity \citep[e.g.,][]{Makishima2001,Brunetti2016}.
Kinetic simulations have shown that reconnection in turbulent plasmas can
rapidly inject particles into a suprathermal population, which may then be
further energized by stochastic interactions with turbulent fluctuations
\citep[e.g.,][]{Comisso2018,Comisso2019}. In the context of cosmological
magnetogenesis, reconnection is particularly relevant because magnetic-field
growth and turbulent current-sheet formation are expected to be closely linked.
The key question is therefore not only whether reconnection is locally faster
than stochastic acceleration, but also whether the magnetic-energy reservoir in
the extremely high-$\beta$ pre-structure plasma is large enough to produce an
energetically meaningful seed population.

In this work, we extend the stochastic-only framework \citep[][]{Ha2026} by adding a
phenomenological reconnection-driven source term to the momentum-space
Fokker--Planck equation for the isotropic ion distribution. The source
normalization is tied to a fraction of the magnetic-energy growth or
dissipation rate associated with pressure-anisotropy-driven magnetogenesis.
This prescription allows us to test whether reconnection can provide seed
particles above the cooling-dominated low-energy regime and whether those
particles can survive cosmological expansion and Coulomb cooling. We also
derive a critical effective reconnection efficiency that determines when the
reconnection injection time becomes shorter than the relevant loss time.

The paper is organized as follows. In Section~\ref{sec:s2}, we
compare the reconnection acceleration time with the stochastic acceleration
time and motivate a two-stage injection--reacceleration picture. In
Section~\ref{sec:s3}, we describe the magnetic-field evolution and
define the reconnection injection power. In Section~\ref{sec:s4}, we
derive the survival and reacceleration criteria, including the critical
effective reconnection efficiency. In Section~\ref{sec:s5}, we
introduce the Fokker--Planck equation with the reconnection source term and
present the resulting particle distributions. In Section~\ref{sec:s6}, we
use a test-particle shock reprocessing model to estimate how the
reconnection-produced seed tail may be reaccelerated by later
structure-formation shocks. Finally, Section~\ref{sec:s7} summarizes the
implications for pre-structure CR seed formation.

\section{Reconnection acceleration timescale}
\label{sec:s2}

Before specifying the reconnection-driven source term, it is useful to
compare the characteristic acceleration time associated with reconnection
with the stochastic acceleration time. This comparison motivates treating
reconnection as a fast injection channel, while stochastic acceleration acts
as a subsequent reacceleration process.

The reconnection acceleration timescale can be estimated from the
reconnection electric field. Writing
\begin{equation}
E_{\rm rec}^{\rm (field)}
\sim
\frac{V_{\rm rec}}{c}B ,
\end{equation}
the energy gain rate of an ion is
\begin{equation}
\dot E_{\rm rec}
\sim
eE_{\rm rec}^{\rm (field)}c
\sim
eBV_{\rm rec},
\end{equation}
where $V_{\rm rec}$ is the reconnection inflow speed. The corresponding
acceleration timescale is therefore
\begin{equation}
t_{\rm rec,acc}(E,z)
\equiv
\frac{E}{\dot E_{\rm rec}}
\sim
\frac{E}{eB(z)V_{\rm rec}(z)} .
\label{eq:t_rec_acc}
\end{equation}

In the optimistic instability-mediated scattering model \citep[][]{Ha2025, Ha2026}, the stochastic acceleration timescale is
estimated as
\begin{equation}
t_{\rm acc}(E,z)
\sim
\left(\frac{c}{V_{\rm tur}(z)}\right)^2
\frac{\lambda(E,z)}{c},
\label{eq:tacc_general}
\end{equation}
where $V_{\rm tur}$ is the characteristic turbulent velocity and
$\lambda(E,z)$ is the particle mean free path. We approximate the mean free
path as
\begin{equation}
\lambda(E,z)\approx \frac{V(E)}{\nu_{\rm eff}(z)} ,
\label{eq:mfp_nueff}
\end{equation}
where $V(E)$ is the particle speed. The effective scattering rate associated
with pressure-anisotropy-driven microinstabilities \citep[e.g.,][]{Falceta2015} is parameterized as
\begin{equation}
\nu_{\rm eff}(z)
\approx
\left(|\Delta|-2\beta^{-1}\right)^{3/2}\omega_{\rm ci}(z),
\label{eq:nueff}
\end{equation}
with the ion gyrofrequency
\begin{equation}
\omega_{\rm ci}(z)=\frac{eB(z)}{m_i c}.
\end{equation}
Here \(\Delta\) is the pressure anisotropy and \(\beta\) is the plasma beta.
This prescription should be interpreted as a thermal-scale scattering closure.
It characterizes the pitch-angle scattering rate induced by
pressure-anisotropy-driven microinstabilities and is most directly applicable
to thermal or near-thermal ions. For particles that have already entered the
suprathermal regime, the effective scattering rate may depend on rigidity and
on the spectrum of magnetic fluctuations, leading to an energy-dependent mean
free path and diffusion coefficient. We therefore do not regard equation~\eqref{eq:nueff} as a universal transport law for high-energy particles,
but use it as an optimistic normalization for estimating the maximum possible
efficiency of instability-mediated stochastic reacceleration. If the scattering
efficiency decreases with rigidity, the reacceleration time would be longer
and the high-energy tail would be further suppressed.

If the reconnection speed is parameterized as
\begin{equation}
V_{\rm rec}(z)=\epsilon_{\rm rec}V_{\rm tur}(z),
\end{equation}
the ratio between the two acceleration timescales becomes
\begin{equation}
\frac{t_{\rm rec,acc}}{t_{\rm acc}}
\sim
\frac{V_{\rm tur}^2}{c\,V_{\rm rec}}
\sim
\frac{V_{\rm tur}(z)}{\epsilon_{\rm rec}c}.
\label{eq:t_ratio_epsrec}
\end{equation}
Since the turbulent velocity in the pre-structure intergalactic medium is
subrelativistic, $V_{\rm tur}\ll c$, reconnection acceleration can be much
faster than stochastic acceleration for plausible values of $\epsilon_{\rm rec}$.
In fast reconnection models and kinetic simulations, the normalized
reconnection rate is often found to be of order $0.01$--$0.1$, with
collisionless reconnection approaching the commonly quoted value
$\sim0.1$ \citep[e.g.,][]{Cassak2017,Liu2017}. We therefore regard
$\epsilon_{\rm rec}\sim0.1$ as an optimistic but standard order-of-magnitude
choice, while smaller values would make reconnection injection correspondingly
slower.
This motivates a two-stage picture:
\begin{equation}
\text{thermal pool}
\;\xrightarrow{\;t_{\rm rec,acc}\;}\;
E\gtrsim E_{\rm C}
\;\xrightarrow{\;t_{\rm acc}\;}\;
\text{suprathermal tail},
\end{equation}
where $E_{\rm C}$ is the Coulomb threshold energy.

\begin{figure}[t]
\centering
\includegraphics[width=0.5\textwidth]{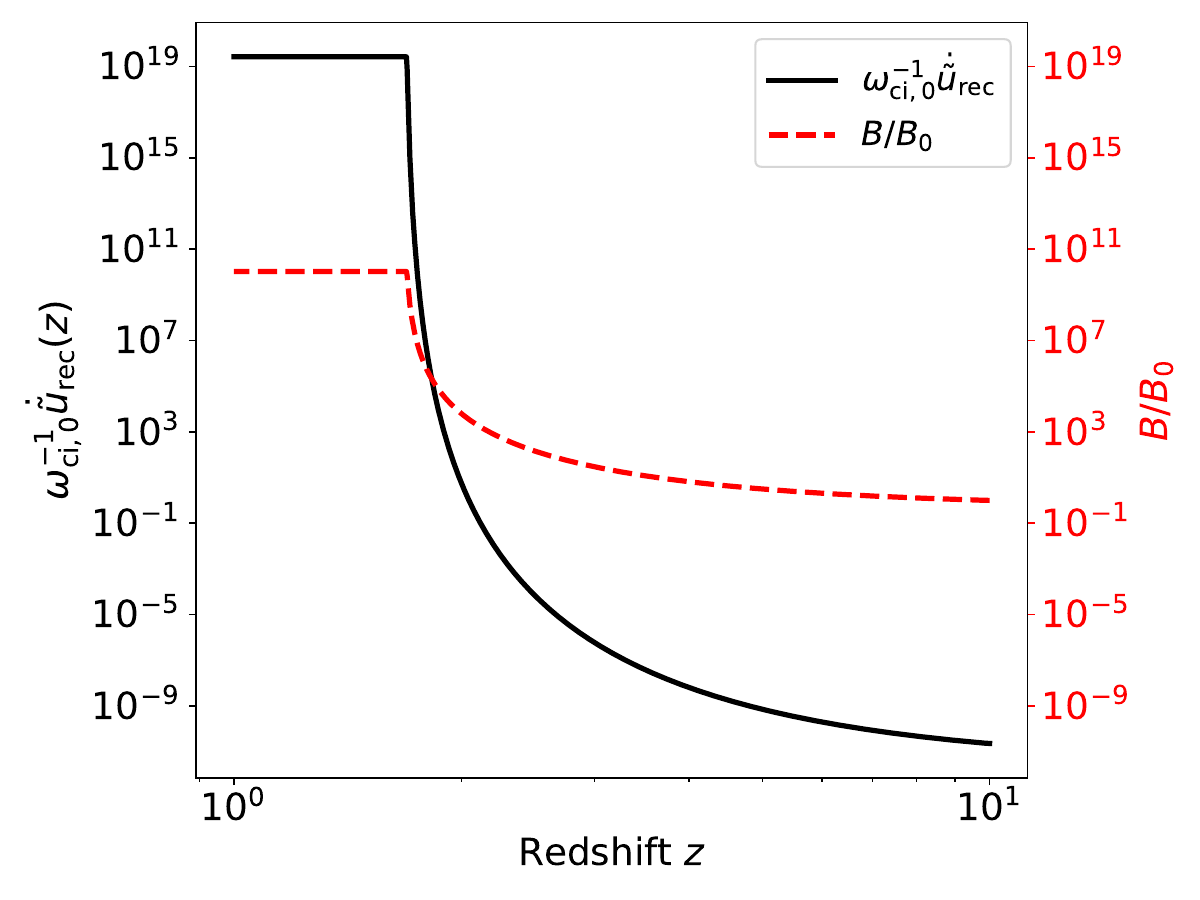}
\caption{
Redshift evolution of the normalized reconnection injection rate
$\omega_{\rm ci,0}^{-1}\dot{\tilde u}_{\rm rec}(z)$ (black solid curve)
for $\eta_{\rm rec}=10^{-8}$ and $|\Delta|-2\beta^{-1} = 0.1|\Delta|$, shown together with the magnetic-field amplification
factor $B/B_0$ (red dashed curve). The right axis uses the same logarithmic
range as the left axis for visual comparison.
}
\label{fig:f1}
\end{figure}

\section{Magnetic-field evolution and reconnection power}
\label{sec:s3}

The magnetic field is assumed to evolve according to an instability-driven magnetogenesis equation of the form
\begin{equation}
\frac{1}{B_0}\frac{dB}{dt}
\approx
\omega_{\rm ci,0}
\frac{
\left(|\Delta|-2\beta^{-1}\right)^{3/2}
}
{1+|\Delta|}
\left(\frac{B}{B_0}\right)^2 ,
\label{eq:B_growth}
\end{equation}
where $B_0$ is the seed magnetic field, $\beta$ is the plasma beta, $\Delta$ is the pressure anisotropy factor and $\omega_{\rm ci,0}={eB_0}/{m_i c}$ is the ion gyrofrequency evaluated at $B_0$ \citep[e.g.,][]{Falceta2015,Ha2025}.
The factor $|\Delta|-2\beta^{-1}$ measures the excess pressure anisotropy
above the marginal instability threshold. Thus, instability-driven growth
requires
\begin{equation}
|\Delta|>2\beta^{-1}.
\end{equation}
We define the supercritical excess anisotropy as
\begin{equation}
\Delta_{\rm ex}
\equiv
|\Delta|-2\beta^{-1}.
\label{eq:Delta_ex_def}
\end{equation}
In the fiducial calculations, we parameterize this excess as
\begin{equation}
\Delta_{\rm ex}
=
f_{\rm ex}|\Delta|,
\qquad
f_{\rm ex}=0.1,
\label{eq:fex_def}
\end{equation}
representing a weakly supercritical plasma in which the pressure anisotropy is
regulated close to marginal stability by instability-induced scattering.
Equivalently, this corresponds to a state in which the anisotropy exceeds the
marginal threshold by a modest fraction of its absolute value. For such a
marginally regulated plasma, $|\Delta|$ remains of order $2\beta^{-1}$, so that
$\Delta_{\rm ex}\propto \beta^{-1}$ for fixed $f_{\rm ex}$.

The reconnection-driven injection power is assumed to be a fraction
$\eta_{\rm rec}$ of the magnetic-energy growth or dissipation rate. We define
\begin{equation}
\dot u_{\rm rec}(z)
\equiv
\eta_{\rm rec}
\left|
\frac{dU_B}{dt}
\right|,
\label{eq:urec_def}
\end{equation}
where
\begin{equation}
U_B(z)=\frac{B^2(z)}{8\pi}
\end{equation}
is the magnetic-energy density. Therefore,
\begin{equation}
\dot u_{\rm rec}(z)
=
\eta_{\rm rec}
\left|
\frac{B(z)}{4\pi}
\frac{dB}{dt}
\right|.
\label{eq:urec_bdbdt}
\end{equation}
For numerical calculations, it is convenient to work with the
magnetic-energy injection rate normalized by $B_0^2$. We therefore define
\begin{equation}
\dot{\tilde u}_{\rm rec}(z)
\equiv
\frac{\dot u_{\rm rec}(z)}{B_0^2}.
\end{equation}
Using equation~\eqref{eq:B_growth}, this normalized injection rate can be written as
\begin{equation}
\dot{\tilde u}_{\rm rec}(z)
=
\eta_{\rm rec}
\left|
\frac{1}{4\pi}
\frac{B}{B_0}
\left(\frac{1}{B_0}\frac{dB}{dt}\right)
\right|
=
\omega_{\rm ci,0}
\frac{\eta_{\rm rec}}{4\pi}
\frac{
\Delta_{\rm ex}^{3/2}
}
{1+|\Delta|}
\left(\frac{B}{B_0}\right)^3 .
\label{eq:urec_growth_dimless}
\end{equation}
This expression links the reconnection-driven injection rate directly to the magnetic-energy growth associated with pressure-anisotropy-driven magnetogenesis.

Here $\eta_{\rm rec}$ is a phenomenological effective efficiency, not a
microscopic reconnection rate. It represents the net fraction of the
magnetic-energy growth or dissipation rate that ultimately appears as a
surviving suprathermal proton source. In the high-$\beta$ pre-structure
plasma, most of the magnetic-energy budget is expected to be associated with
field amplification, turbulent plasma motions, and thermal heating, while only
a small fraction is converted into nonthermal ion energization. We therefore
regard $\eta_{\rm rec}\ll1$ as a natural volume-averaged efficiency for
reconnection-driven proton seed formation. Determining this factor from first
principles would require kinetic simulations of reconnection in high-$\beta$
cosmological plasmas and is beyond the scope of this work.

Figure~\ref{fig:f1} illustrates the redshift evolution of the normalized reconnection injection rate for a representative value $\eta_{\rm rec}=10^{-8}$, together with the magnetic-field amplification factor $B/B_0$.
Because the normalized injection rate scales as $\dot{\tilde u}_{\rm rec}\propto (B/B_0)^3$, it rises much more rapidly than the magnetic-field amplitude during the nonlinear magnetogenesis phase.
This behavior indicates that reconnection-driven particle injection is expected to be strongly concentrated near the epoch of rapid magnetic-field amplification, rather than being uniformly distributed over the entire pre-structure interval.

\section{Survival and reacceleration criteria}
\label{sec:s4}

The first relevant loss timescale is the cosmological expansion time, which we
identify with the Hubble time,
\begin{equation}
t_H(z)=H^{-1}(z),
\end{equation}
where
\begin{equation}
H(z)=H_0\sqrt{\Omega_m(1+z)^3+\Omega_\Lambda}.
\label{eq:Hz}
\end{equation}
Here $H_0$ is the present-day Hubble constant, and $\Omega_m$ and
$\Omega_\Lambda$ are the present-day matter and dark-energy density
parameters, respectively. In the numerical calculations, we adopt rounded fiducial values for a flat
$\Lambda$CDM cosmology with $H_0=70\,{\rm km\,s^{-1}\,Mpc^{-1}}$, $\Omega_m=0.3$, and
$\Omega_\Lambda=0.7$ \citep[e.g.,][]{Komatsu2011,Planck2020}.

The second relevant loss timescale is the Coulomb cooling time. We define
\begin{equation}
t_{\rm C}(E,z)
=
\frac{E}{|\dot E_{\rm C}(E,z)|}.
\label{eq:tC_def}
\end{equation}
For non-relativistic or mildly relativistic ions interacting with thermal
electrons in a fully ionized plasma, we approximate the Coulomb energy-loss
rate as
\begin{equation}
|\dot E_{\rm C}(E,z)|
\approx
\frac{4\pi e^4 n_e(z)\ln\Lambda}{m_e V(E)} ,
\label{eq:Edot_C}
\end{equation}
so that
\begin{equation}
t_{\rm C}(E,z)
\approx
\frac{E\,m_e V(E)}
{4\pi e^4 n_e(z)\ln\Lambda}.
\label{eq:tC_explicit}
\end{equation}
Here $V(E)$ is the ion speed, $n_e(z)$ is the free-electron number density,
and $\ln\Lambda$ is the Coulomb logarithm.
In the fiducial calculation, the electron density is taken to scale with
cosmic expansion as
\begin{equation}
n_e(z)=n_{e,0}(1+z)^3 ,
\end{equation}
appropriate for a fully ionized intergalactic medium.

We then define the combined loss time as
\begin{equation}
t_{\rm loss}^{-1}(E,z)
=
t_H^{-1}(z)+t_{\rm C}^{-1}(E,z).
\label{eq:tloss_def}
\end{equation}
This definition reduces to the shorter of the two timescales in the limits
$t_{\rm C}\ll t_H$ or $t_H\ll t_{\rm C}$.
A necessary condition for reconnection-injected particles to survive losses is
\begin{equation}
t_{\rm loss}(E_{\rm rec},z)
>
t_{\rm rec,inj}(z),
\label{eq:survival_condition}
\end{equation}
where \(t_{\rm rec,inj}\) is the characteristic injection timescale. The characteristic reconnection injection time is estimated from the inverse of the normalized reconnection power. Using $\dot{\tilde u}_{\rm rec}=\dot u_{\rm rec}/B_0^2$, we define
\begin{equation}
\omega_{\rm ci,0}t_{\rm rec,inj}(z)
\sim
\left[
\omega_{\rm ci,0}^{-1}\dot{\tilde u}_{\rm rec}(z)
\right]^{-1}.
\label{eq:trecinj_dimless}
\end{equation}
This quantity should be interpreted as a normalized injection timescale rather
than as the time required to build a prescribed, redshift-independent seed
energy density. It is used as a diagnostic measure of how rapidly
reconnection-driven injection can operate relative to the relevant loss time.

Once particles are injected into the suprathermal regime, subsequent stochastic
reacceleration is possible only if
\begin{equation}
t_{\rm acc}(E,z)
<
t_{\rm loss}(E,z).
\label{eq:reacc_condition}
\end{equation}
This condition requires the instability-mediated acceleration time to be
shorter than the combined loss time due to cosmological expansion and Coulomb
cooling.

\begin{figure}[t]
\centering
\includegraphics[width=0.45\textwidth]{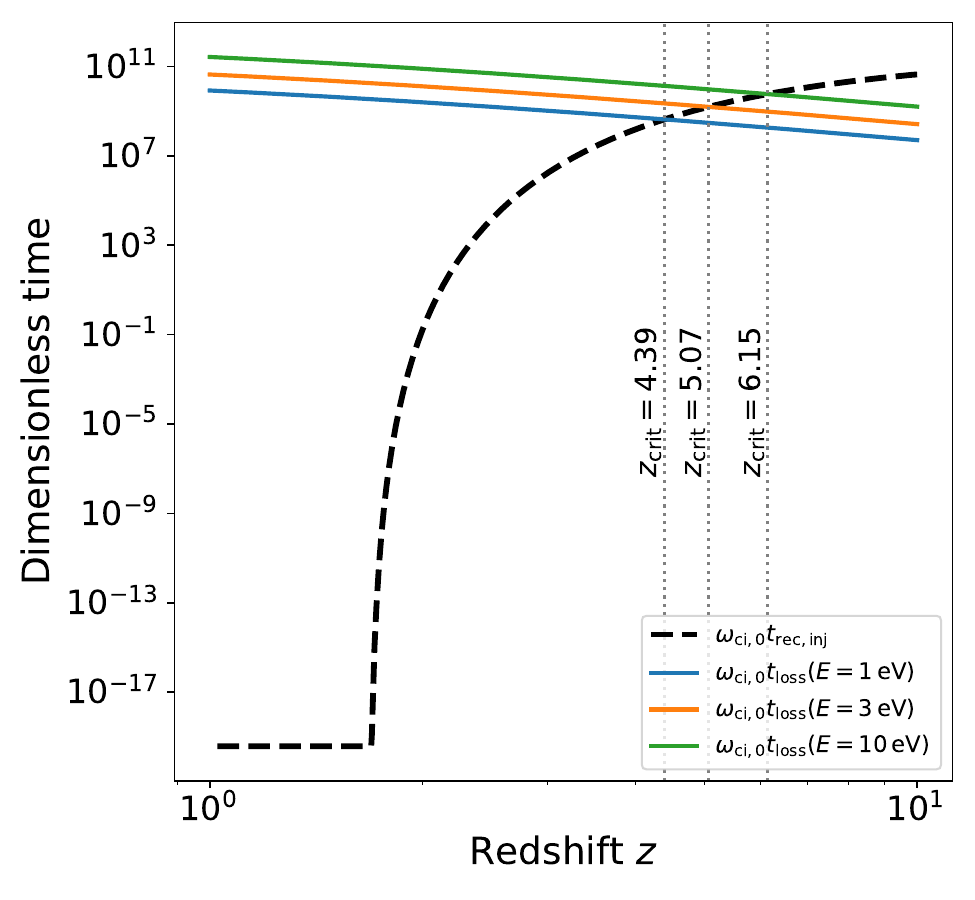}
\caption{
Comparison of the dimensionless loss time
$\omega_{\rm ci,0}t_{\rm loss}$ for different injected particle energies with the
dimensionless reconnection injection time
$\omega_{\rm ci,0}t_{\rm rec,inj}$ for $\eta_{\rm rec}=10^{-8}$ and $\Delta_{\rm ex} = 0.1|\Delta|$.
Vertical dotted lines mark the critical redshifts where
$t_{\rm loss}=t_{\rm rec,inj}$.
}
\label{fig:f2}
\end{figure}

\begin{figure}[t]
\centering
\includegraphics[width=0.5\textwidth]{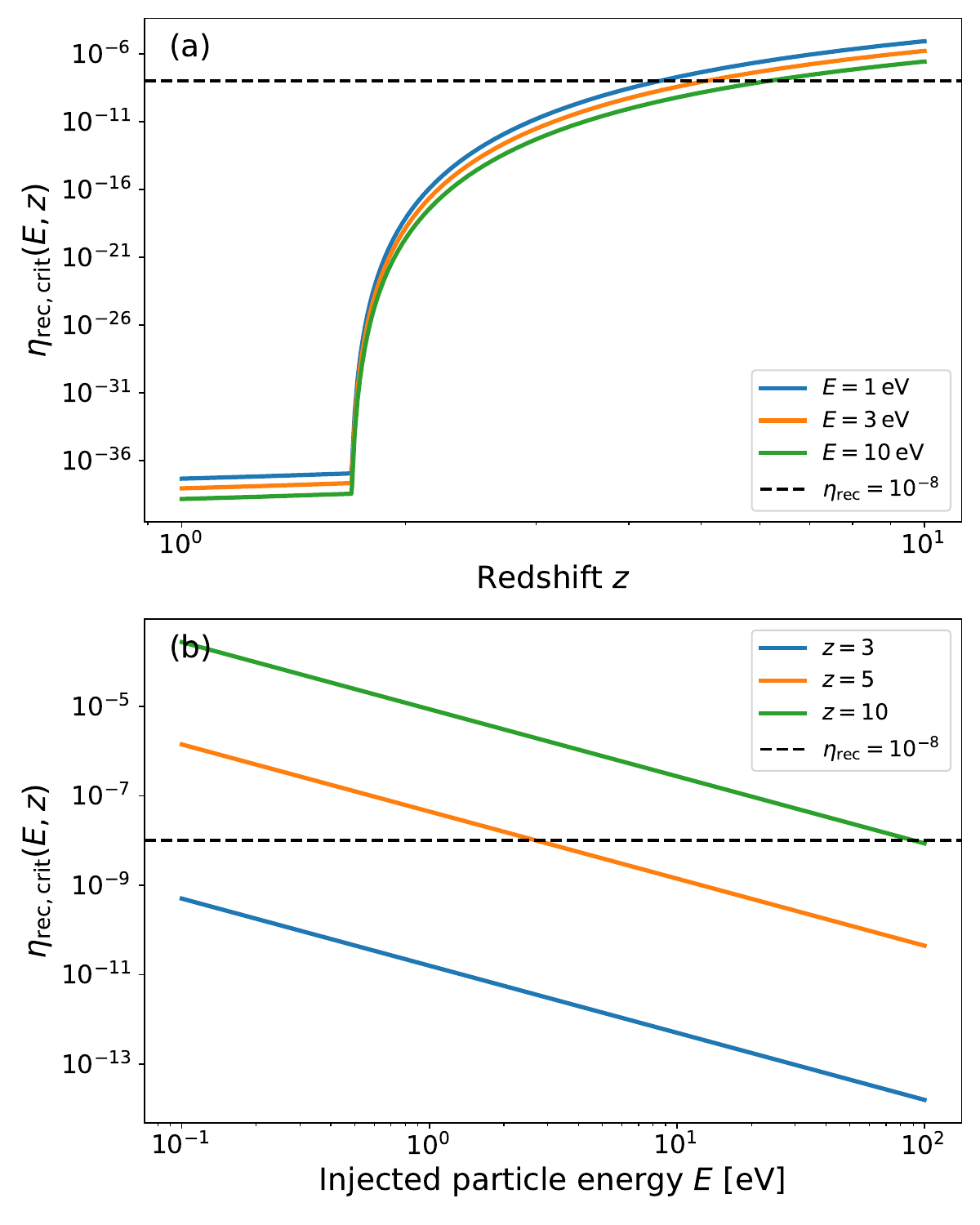}
\caption{
Critical effective reconnection efficiency $\eta_{\rm rec,crit}(E,z)$.
Panel (a) shows $\eta_{\rm rec,crit}$ as a function of redshift for fixed
injected particle energies $E=1$, 3, and 10 eV. Panel (b) shows
$\eta_{\rm rec,crit}$ as a function of injected particle energy at fixed
redshifts $z=3$, 5, and 10. The dashed horizontal line marks
$\eta_{\rm rec}=10^{-8}$. $\Delta_{\rm ex} = 0.1|\Delta|$ is used for all calculations.
}
\label{fig:f3}
\end{figure}

Figure~\ref{fig:f2} compares the dimensionless loss time,
$\omega_{\rm ci,0}t_{\rm loss}$, with the reconnection injection time,
$\omega_{\rm ci,0}t_{\rm rec,inj}$, for a representative effective injection
efficiency $\eta_{\rm rec}=10^{-8}$. The comparison illustrates that the
competition between reconnection injection and Coulomb cooling is strongly
energy dependent. For the parameters shown here, particles with energies of
$E=1$, 3, and 10 eV cross the condition
$t_{\rm C}=t_{\rm rec,inj}$ at progressively higher redshifts,
$z_{\rm crit}\approx 4.39$, 5.07, and 6.15, respectively. At redshifts below
the corresponding $z_{\rm crit}$, the reconnection injection time becomes
shorter than the energy loss time, so that injected particles can survive
cooling more efficiently. This suggests that even for very small effective
values of $\eta_{\rm rec}$, reconnection-driven injection can become viable
during the later stages of magnetogenesis, with the onset redshift depending
sensitively on the injected particle energy.

The choice $\eta_{\rm rec}=10^{-8}$ should be understood as a conservative
effective efficiency. For this value, reconnection-driven injection already
becomes viable at $z_{\rm crit}\sim 5$--6 for the particle energies shown in
Figure~\ref{fig:f2}. Increasing $\eta_{\rm rec}$ would move the crossing to
higher redshifts, possibly approaching $z\sim 10$, but at such early epochs the
magnetic field is still weak and the available magnetic-energy reservoir is
small. Thus, a larger effective efficiency would not by itself guarantee a
dynamically important seed population. In the following, we therefore use
$\eta_{\rm rec}=10^{-8}$ as a fiducial value that probes the regime where
reconnection injection becomes possible during the late, more strongly
magnetized phase of the evolution.

The critical effective reconnection efficiency can be obtained by imposing
$t_{\rm rec,inj}=t_{\rm loss}$. Using $t_{\rm rec,inj}\sim \dot{\tilde u}_{\rm rec}^{-1}$, we obtain the critical reconnection efficiency $\eta_{\rm rec,crit}$
\begin{equation}
\eta_{\rm rec,crit}(E,z)
=
4\pi
\frac{1+|\Delta|}
{\Delta_{\rm ex}^{3/2}}
\left(\frac{B}{B_0}\right)^{-3}
\frac{1}{\omega_{\rm ci,0}t_{\rm loss}(E,z)} .
\label{eq:eta_rec_crit}
\end{equation}
For $\eta_{\rm rec}>\eta_{\rm rec,crit}$, reconnection injection proceeds
faster than the relevant loss processes, allowing injected suprathermal
particles to survive. For $\eta_{\rm rec}<\eta_{\rm rec,crit}$, losses
dominate before a significant seed population can be established.

Figure~\ref{fig:f3} shows the critical effective reconnection efficiency
$\eta_{\rm rec,crit}(E,z)$ obtained from the condition
$t_{\rm rec,inj}=t_{\rm loss}$. Panel (a) shows the redshift dependence for
fixed injected particle energies, while panel (b) shows the energy dependence
at representative redshifts. The dashed horizontal line indicates
$\eta_{\rm rec}=10^{-8}$, used as a fiducial effective injection efficiency.
For $\eta_{\rm rec}>\eta_{\rm rec,crit}$, reconnection injection is faster than
the relevant loss processes and injected particles can survive, whereas for
$\eta_{\rm rec}<\eta_{\rm rec,crit}$ losses dominate. The strong decrease of
$\eta_{\rm rec,crit}$ toward lower redshift reflects the rapid growth of the
magnetic field and the corresponding increase of the reconnection injection
power. The energy dependence indicates that higher-energy injected particles
require a smaller effective reconnection efficiency to survive Coulomb cooling,
consistent with the longer Coulomb loss time at higher energies.

\section{Transport equation and reconnection-driven injection}
\label{sec:s5}

We follow the momentum-space Fokker--Planck framework developed in
\cite{Ha2026} and extend it by adding an explicit reconnection-driven source
term.
We consider the isotropic phase-space distribution of ions, $f(p,z)$, normalized such that
\begin{equation}
dn = 4\pi p^2 f(p,z)\,dp .
\end{equation}
Neglecting spatial gradients, the momentum-space Fokker--Planck equation is
\begin{equation}
\frac{\partial f}{\partial t}
=
\frac{1}{p^2}
\frac{\partial}{\partial p}
\left[
p^2 D_{pp}(p,z)
\frac{\partial f}{\partial p}
\right]
-
\frac{1}{p^2}
\frac{\partial}{\partial p}
\left[
p^2 \dot p_{\rm loss}(p,z) f
\right]
+
Q_{\rm rec}(p,z),
\label{eq:fp_t}
\end{equation}
where $D_{pp}$ is the stochastic momentum diffusion coefficient, $\dot p_{\rm loss}$ is the systematic momentum-loss rate, and $Q_{\rm rec}$ is the reconnection-driven injection term.
Cosmological evolution is incorporated through
\begin{equation}
\frac{dt}{dz}
=
-\frac{1}{(1+z)H(z)},
\label{eq:dtdz}
\end{equation}
where $H(z)$ is defined in equation~\eqref{eq:Hz}.
The transport equation can therefore be written as
\begin{align}
-(1+z)H(z)\frac{\partial f}{\partial z}
&=
\frac{1}{p^2}
\frac{\partial}{\partial p}
\left[
p^2 D_{pp}(p,z)
\frac{\partial f}{\partial p}
\right]
\nonumber\\
&-
\frac{1}{p^2}
\frac{\partial}{\partial p}
\left[
p^2 \dot p_{\rm loss}(p,z) f
\right]
+
Q_{\rm rec}(p,z).
\label{eq:fp_z}
\end{align}

For the stochastic reacceleration term, we adopt the minimal closure
\begin{equation}
D_{pp}(p,z)=\frac{p^2}{t_{\rm acc}(p,z)} ,
\label{eq:Dpp_general}
\end{equation}
where $t_{\rm acc}$ is the stochastic acceleration time defined in equation \eqref{eq:tacc_general}.
The loss term includes adiabatic momentum losses due to cosmological expansion and Coulomb losses,
\begin{equation}
\dot p_{\rm loss}(p,z)
=
\dot p_{\rm ad}(p,z)
+
\dot p_{\rm C}(p,z).
\label{eq:pdot_loss}
\end{equation}
The adiabatic term is
\begin{equation}
\dot p_{\rm ad}(p,z)
=
-H(z)p,
\label{eq:pdot_ad}
\end{equation}
and the Coulomb momentum-loss rate can be approximated as
\begin{equation}
\dot p_{\rm C}(p,z)
\approx
-
\frac{4\pi e^4 n_e(z)\ln\Lambda}
{m_e v^2(p)} .
\label{eq:pdot_coul}
\end{equation}

To solve the Fokker--Planck equation, we initialize the ion distribution at
$z=10$ as a non-relativistic Maxwellian,
\begin{equation}
f_{\rm M}(p,z=10)
=
\frac{n_i}{(2\pi m_i k_B T_0)^{3/2}}
\exp\left[
-\frac{p^2}{2m_i k_B T_0}
\right],
\label{eq:initial_maxwellian}
\end{equation}
where $n_i$ is the ion number density. This initial condition is not intended
to represent the volume-averaged thermal state of the Universe at $z=10$.
Rather, it describes a simplified diffuse ionized medium used as a fiducial
environment for assessing the competition among reconnection-driven injection,
stochastic reacceleration, cosmological expansion, and Coulomb cooling.
Accordingly, $T_0$ should be interpreted as an effective model parameter. In
the fiducial calculation, we adopt $E_{\rm th} \sim k_B T_0 = 1\,{\rm eV}$.

\begin{figure}[t]
\centering
\includegraphics[width=0.5\textwidth]{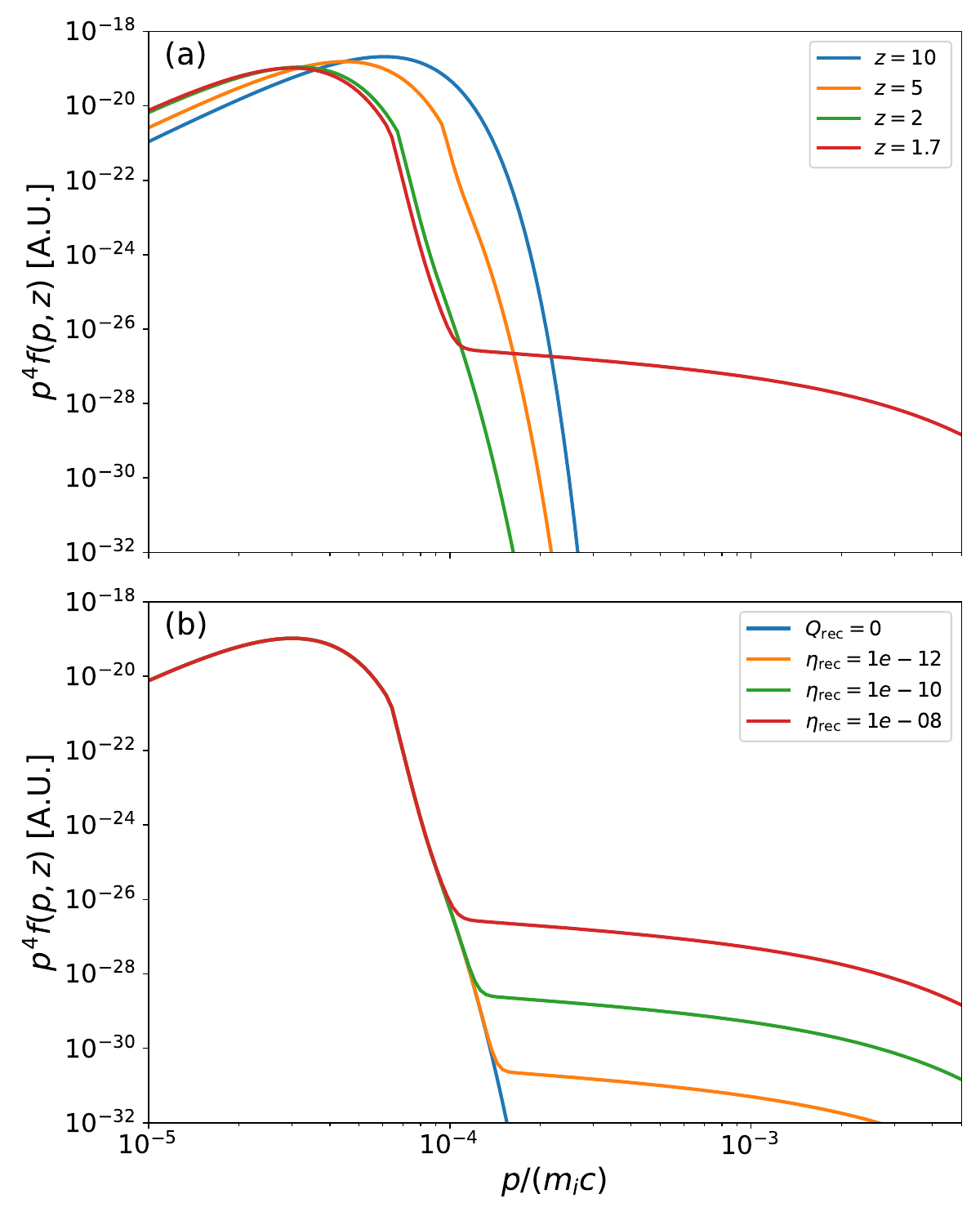}
\caption{
Fokker--Planck solutions with reconnection-driven injection. Panel (a) shows the redshift evolution of $p^4 f(p,z)$ for the fiducial case $\eta_{\rm rec}=10^{-8}$. Panel (b) compares the distributions at $z=1.7$ for $Q_{\rm rec}=0$ and for $\eta_{\rm rec}=10^{-12},10^{-10},10^{-8}$. For all cases, $s_{\rm rec} = 4.5$ is used.
}
\label{fig:f4}
\end{figure}

We model reconnection-driven injection as a phenomenological power-law source
with an exponential cutoff,
\begin{equation}
Q_{\rm rec}(p,z)
=
A_{\rm rec}(z)
p^{-s_{\rm rec}}
\exp\left[-\frac{p}{p_{\rm cut}(z)}\right]
H[p-p_{\rm rec}(z)] ,
\label{eq:Qrec_powerlaw}
\end{equation}
where $s_{\rm rec}$ is the injection spectral index, $p_{\rm rec}$ is the
minimum injection momentum, $p_{\rm cut}$ is the cutoff momentum, and $H$ is
the Heaviside function. The normalization is fixed by requiring that the
injected suprathermal energy density per unit time equals the reconnection
power defined in equation~\eqref{eq:urec_def},
\begin{equation}
\int_0^\infty 4\pi p^2 E(p)Q_{\rm rec}(p,z)\,dp
=
\dot u_{\rm rec}(z).
\label{eq:Qrec_norm}
\end{equation}
This gives
\begin{equation}
A_{\rm rec}(z)
=
\frac{\dot u_{\rm rec}(z)}
{4\pi
\displaystyle\int_{p_{\rm rec}}^\infty
p^{2-s_{\rm rec}}E(p)
\exp[-p/p_{\rm cut}(z)]\,dp } .
\label{eq:Arec}
\end{equation}

In the fiducial calculation, we adopt $s_{\rm rec}=4.5$, corresponding to a
steep suprathermal injection spectrum. This should be regarded as a
phenomenological source-shape parameter rather than a unique prediction of
reconnection theory. Because the total normalization of $Q_{\rm rec}$ is fixed
by the magnetic-energy growth rate, variations in $s_{\rm rec}$ mainly
redistribute the injected power in momentum space. The conclusion that the
reconnection-produced tail is energetically subdominant is therefore controlled
primarily by the high-$\beta$ magnetic-energy budget rather than by the precise
choice of $s_{\rm rec}$.

The minimum injection momentum and the cutoff momentum are determined from the
corresponding kinetic energies as
\begin{equation}
p_{\rm rec}(z)
=
\frac{1}{c}
\sqrt{
E_{\rm rec}(z)
\left[
E_{\rm rec}(z)+2m_i c^2
\right]
},
\label{eq:prec}
\end{equation}
and
\begin{equation}
p_{\rm cut}(z)
=
\frac{1}{c}
\sqrt{
E_{\rm cut}(z)
\left[
E_{\rm cut}(z)+2m_i c^2
\right]
}.
\label{eq:pcut}
\end{equation}
In the fiducial calculations, these energies are specified relative to the
thermal energy scale. We adopt
\begin{equation}
E_{\rm rec}\sim3E_{\rm th},
\qquad
E_{\rm cut}\sim10^3E_{\rm th},
\end{equation}
corresponding to $E_{\rm rec}\sim 3\,{\rm eV}$ and
$E_{\rm cut}\sim 1\,{\rm keV}$ for the representative thermal scale used in
the numerical calculation. Thus, $p_{\rm rec}$ sets the low-energy threshold of the
reconnection-injected component, while $p_{\rm cut}$ provides a
phenomenological upper scale for the injected spectrum. Whether particles
injected at $p_{\rm rec}$ can survive cooling is determined separately by the
timescale criterion in Figure~\ref{fig:f2}. For the fiducial
$\eta_{\rm rec}=10^{-8}$, the adopted injection energy
$E_{\rm rec}\sim 3\,{\rm eV}$ becomes viable below
$z_{\rm crit}\sim 5.07$, indicating that reconnection injection at this
scale is effective only during the later, more strongly magnetized phase.
The cutoff energy is treated as a phenomenological upper scale of the injected
source spectrum, not as a self-consistent loss-limited maximum energy. This
does not introduce an additional energy budget, because the normalization of
$Q_{\rm rec}$ is fixed by equation~\eqref{eq:Qrec_norm}; varying $E_{\rm cut}$
redistributes the injected power in momentum space while keeping the total
injected energy tied to $\dot u_{\rm rec}$. We adopt
$E_{\rm cut}\sim 1\,{\rm keV}$, far below the maximum ion energy
$\mathcal{O} (10-10^2)\,{\rm GeV}$ estimated for optimistic stochastic acceleration
over a Hubble time in the stochastic-only magnetogenesis model. This choice therefore represents a conservative low-energy
cutoff for the reconnection-injected component.

Figure~\ref{fig:f4} shows the Fokker--Planck solutions including the reconnection-driven source term. Panel (a) presents the redshift evolution of the distribution for the fiducial case $\eta_{\rm rec}=10^{-8}$. As the system evolves toward lower redshift, the thermal component shifts to lower momenta because of cosmological expansion and cooling. Although Figure~\ref{fig:f2} shows that reconnection-driven injection becomes faster than losses for the fiducial injection energy after $z\sim5$, the magnetic-energy reservoir remains small at this stage. The suprathermal component therefore grows only gradually, and cooling continues to affect the distribution until $z\sim2$. A visible high-momentum tail develops mainly during the later, more strongly magnetized phase, when the reconnection power has increased sufficiently.
Panel (b) compares the distributions at $z=1.7$ for different effective reconnection efficiencies. In the absence of reconnection injection $(Q_{\rm rec}=0)$, the high-momentum tail is strongly suppressed. Increasing $\eta_{\rm rec}$ systematically enhances the suprathermal component at
$p/(m_i c)\gtrsim 10^{-4}$, demonstrating that reconnection-driven injection can supply seed particles that are not produced by stochastic acceleration alone. The result shows that the survival and normalization of the seed population are controlled primarily by the effective reconnection efficiency.

Although the reconnection source produces a visible suprathermal tail in the
Fokker--Planck solution, its energetic importance remains limited. To quantify
the energetic importance of the reconnection-produced tail, we define the
suprathermal energy density as
\begin{equation}
u_{\rm nth}(z)
=
\int_{p_{\rm rec}}^{\infty}
4\pi p^2 E(p) f(p,z)\,dp ,
\label{eq:unth_tail}
\end{equation}
where $p_{\rm rec}$ is the minimum momentum of the reconnection-injected
particles. For the fiducial case $\eta_{\rm rec}=10^{-8}$, we find
$u_{\rm nth}/u_{\rm th}\sim 10^{-7}$ at the end of the calculation.
This small energy fraction also implies a negligible nonthermal pressure
contribution. For an order-of-magnitude estimate, the suprathermal pressure may
be written as $P_{\rm nth}\sim (2/3)u_{\rm nth}$ for a non-relativistic tail,
whereas the thermal pressure is $P_{\rm th}\sim (2/3)u_{\rm th}$. Thus,
\begin{equation}
\frac{P_{\rm nth}}{P_{\rm th}}
\sim
\frac{u_{\rm nth}}{u_{\rm th}}
\sim
10^{-7}.
\end{equation}
Even if the suprathermal particles become mildly relativistic, the numerical
prefactor changes only by a factor of order unity and the conclusion remains
unchanged. The reconnection-produced component therefore cannot significantly
modify the thermodynamics of the pre-structure intergalactic medium, the
magnetogenesis history, or the early feedback energy budget.
This value is also many orders of magnitude below the typical fraction of
kinetic-energy flux converted into cosmic rays at structure-formation shocks,
which is commonly expected to lie in the range $\sim 10^{-3}$--$10^{-1}$,
depending on the Mach number and microphysical injection efficiency. Thus,
reconnection-driven injection during cosmological magnetogenesis can generate a
suprathermal seed tail, but it is unlikely to make a dynamically important
contribution to the cosmic-ray energy budget before nonlinear structure
formation.

This limitation follows naturally from the extremely high-$\beta$ nature of
the pre-structure plasma. In the present model, the reconnection source is
normalized to the magnetic-energy growth rate,
$\dot u_{\rm rec}\propto \eta_{\rm rec}|dU_B/dt|$. Since
$U_B=P_{\rm th}/\beta$, the available magnetic-energy reservoir is small when
$\beta\gg1$. Moreover, using the weakly supercritical prescription
$\Delta_{\rm ex}=f_{\rm ex}|\Delta|$ with $|\Delta|\sim 2\beta^{-1}$, the
instability-driving factor scales as
$\Delta_{\rm ex}^{3/2}\propto\beta^{-3/2}$. At fixed thermal pressure,
$\omega_{\rm ci}\propto B\propto\beta^{-1/2}$, and therefore the reconnection
power scales approximately as
\begin{equation}
\dot u_{\rm rec}
\sim
\eta_{\rm rec}U_B\omega_{\rm ci}\Delta_{\rm ex}^{3/2}
\propto
\eta_{\rm rec}\beta^{-3}.
\end{equation}
Thus, reconnection-driven injection is inefficient in the high-$\beta$ limit,
even when the local reconnection acceleration time is shorter than both the
stochastic acceleration time and the relevant loss time.

Consequently, the total energy that can be transferred to suprathermal
particles remains limited by the small magnetic-energy fraction. The resulting
tail may therefore serve as a low-level seed population for subsequent shock
acceleration, but the dominant cosmic-ray energy production is expected to
occur later at structure-formation shocks.

\begin{figure}
\centering
\includegraphics[width=0.5\textwidth]{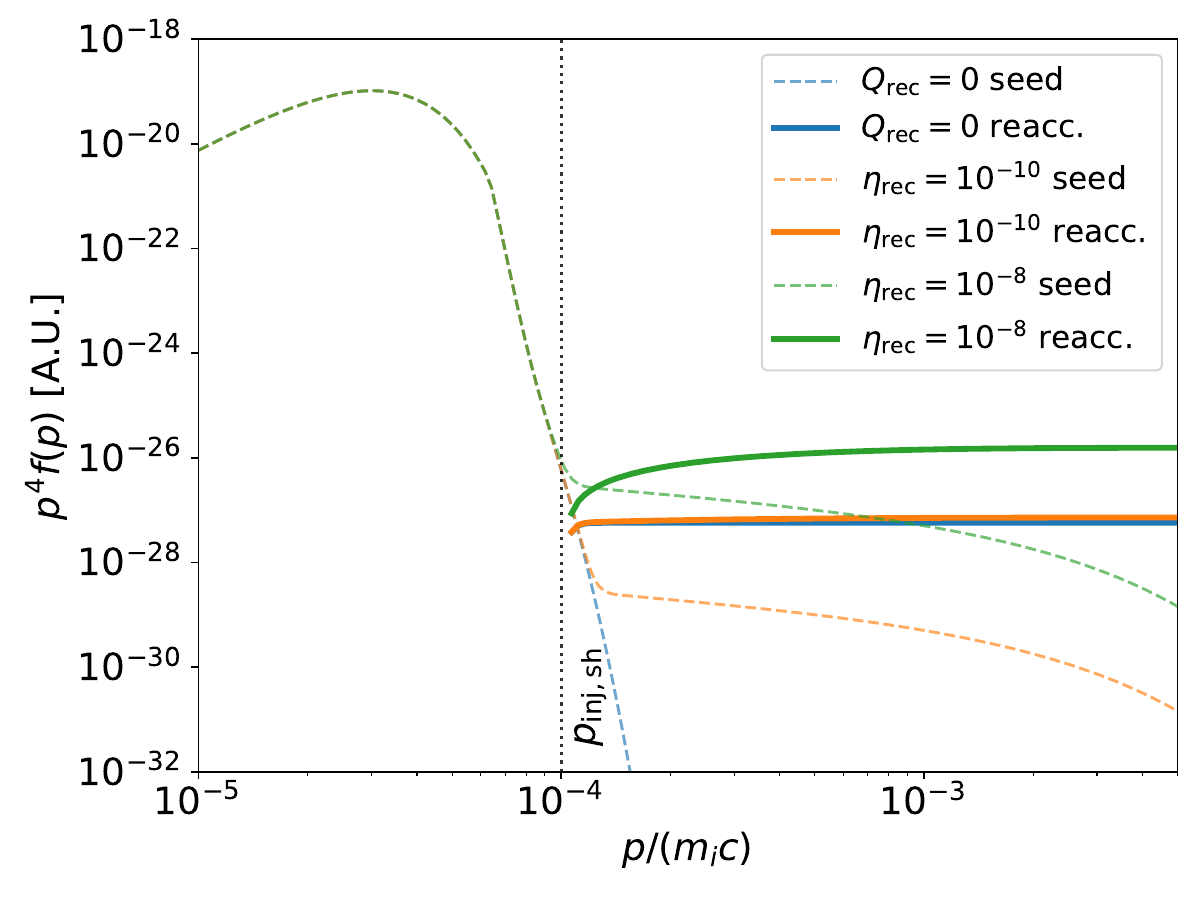}
\caption{
Shock reacceleration of reconnection-produced seed particles. Dashed curves
show the pre-shock seed distributions obtained from the Fokker--Planck
calculation at $z=1.7$, while solid curves show the corresponding
test-particle DSA-reprocessed spectra for $q=4$. The vertical dotted line
marks the adopted shock injection momentum,
$p_{\rm inj,sh}/(m_i c)=10^{-4}$. Reconnection-driven injection enhances the
suprathermal seed population available for subsequent shock acceleration.
}
\label{fig:f5}
\end{figure}

\section{Shock reacceleration of reconnection-produced seeds}
\label{sec:s6}

The results above show that reconnection-driven injection can produce a
suprathermal tail, although the associated nonthermal energy fraction remains
small. This suggests that reconnection during cosmological magnetogenesis is
unlikely to be the dominant source of cosmic-ray energy by itself. However,
the reconnection-produced tail may still be relevant as a seed population for
subsequent diffusive shock acceleration once structure-formation shocks emerge.
To assess this possibility, we apply a simple test-particle shock
reacceleration model as a post-processing diagnostic.

For a pre-existing upstream seed distribution $f_{\rm seed}(p)$, the
test-particle diffusive shock acceleration solution \citep[e.g.,][]{Kang2011} can be written as
\begin{equation}
f_{\rm sh}(p)
=
q p^{-q}
\int_{p_{\rm inj,sh}}^{p}
p'^{\,q-1} f_{\rm seed}(p')\,dp' ,
\label{eq:shock_reacc}
\end{equation}
where $p_{\rm inj,sh}$ is the minimum momentum of particles participating in
shock acceleration. The test-particle momentum slope is
\begin{equation}
q=\frac{3r}{r-1},
\label{eq:q_shock}
\end{equation}
where $r$ is the shock compression ratio. For a strong non-relativistic shock,
$r\approx4$ and hence $q\approx4$. In the following, we adopt $q=4$, corresponding to the strong-shock
test-particle limit, and take
$p_{\rm inj,sh}=10^{-4}m_i c \sim 3p_{\rm th}$ as a representative shock
injection threshold. This value should not be interpreted as a universal
injection momentum, but as a standard order-of-magnitude scale motivated by
kinetic simulations, where efficient
entry into diffusive shock acceleration typically requires pre-energization to
a few times the downstream thermal proton momentum
\citep[e.g.,][]{Caprioli2014,Park2015,Ha2018b,Ryu2019}.

We identify $f_{\rm seed}(p)$ with the distributions obtained from the
Fokker--Planck calculation at $z=1.7$. This choice is intended to represent
the seed population available shortly before the onset of efficient
structure-formation shock acceleration. We compare three cases:
$Q_{\rm rec}=0$, $\eta_{\rm rec}=10^{-10}$, and
$\eta_{\rm rec}=10^{-8}$. The first case provides the stochastic-only baseline,
whereas the latter two cases quantify how reconnection-produced suprathermal
particles modify the seed population available to the shock.

Figure~\ref{fig:f5} shows the result of this shock reprocessing calculation.
The dashed curves show the pre-shock seed distributions, while the solid curves
show the corresponding shock-reprocessed spectra. In the absence of
reconnection injection, the seed population above $p_{\rm inj,sh}$ is strongly
suppressed, and the resulting reprocessed component remains weak. When
reconnection-driven injection is included, the pre-existing suprathermal tail
provides additional particles above the shock injection threshold. As a result,
the shock-reprocessed component is enhanced relative to the $Q_{\rm rec}=0$
case. For the fiducial $\eta_{\rm rec}=10^{-8}$ model, the suprathermal proton
population above $p_{\rm inj,sh}$ is enhanced by roughly an order of magnitude.

This calculation should be interpreted as a diagnostic estimate rather than a
self-consistent model of shock acceleration. It does not include nonlinear
shock modification, self-generated turbulence, particle escape, or the
time-dependent coupling between the shock and the upstream distribution.
Nevertheless, it illustrates an important physical implication of the present
model: reconnection-driven injection may be energetically negligible before
shock formation, but it can still be kinetically relevant by supplying seed
particles that are later reaccelerated by structure-formation shocks.

The reconnection-produced proton tail may also have indirect implications for
hadronic gamma-ray emission after subsequent shock reacceleration. Inelastic
proton--proton interactions generate neutral pions, whose decay produces
gamma-rays. However, the outer regions of galaxy clusters are characterized by
low gas densities, and previous studies have shown that the corresponding
$\pi^0$-decay emission from accretion-shock regions is generally weaker than
the inverse-Compton gamma-ray component and much more difficult to detect
\citep[e.g.,][]{Keshet2003,Miniati2003,Pinzke2010,Ha2023}. In particular, \cite{Ha2023} found that the accretion-shock contribution
to cluster gamma-ray emission is already far below current Fermi-LAT limits
because of the low target gas density in cluster outskirts. Thus, even if a
reconnection-produced seed tail enhances the number of suprathermal protons
available for shock reacceleration by about an order of magnitude, the
resulting hadronic gamma-ray flux from cluster outskirts is still expected to
remain a few orders of magnitude below current Fermi-LAT upper limits \citep[e.g.,][]{Ackermann2014}.
This implies that reconnection-driven seed formation is unlikely to provide a
directly detectable hadronic gamma-ray signature in the pre-structure or
cluster-outskirt environment. Its more plausible role is to modify the seed
population entering later structure-formation shocks, while the observable
gamma-ray output remains controlled primarily by the shock acceleration
efficiency, the accumulated CR proton energy density, and the density of the
target gas.

Previous studies that model proton acceleration at intracluster shocks with
more detailed injection and acceleration physics generally find gamma-ray
fluxes below current Fermi-LAT upper limits \citep[e.g.,][]{Ryu2019,Ha2020}.
One possible additional channel is that CR protons accelerated at outer
accretion shocks could propagate into denser cluster regions and be
reaccelerated by intracluster shocks, thereby increasing the hadronic
gamma-ray output. However, such transport is inefficient on cluster
timescales. A 10 GeV proton population
would require an effective diffusion coefficient of order
$D\sim 10^{32}\,{\rm cm^2\,s^{-1}}$ to travel
$L\sim5\,{\rm Mpc}$ within a cluster age $t\sim5\,{\rm Gyr}$. 
Such rapid transport is optimistic for magnetized intracluster plasmas,
because cluster-scale CR transport generally requires very large effective
diffusion coefficients, of order $10^{31}$--$10^{32}\,{\rm cm^2\,s^{-1}}$,
depending on the assumed magnetic-field topology and transport model
\citep[e.g.,][]{Ensslin2011,Wiener2019}. Similarly, if CRs stream at roughly
the Alfv\'en speed, as expected in the self-confined streaming limit, then
$v_A\sim10^2\,{\rm km\,s^{-1}}$ gives a streaming distance of only
$\sim0.5\,{\rm Mpc}$ over $5\,{\rm Gyr}$
\citep[e.g.,][]{Wiener2013,Wiener2018}. Therefore, the reconnection-seeded
protons reaccelerated at accretion shocks are expected to interact mainly in
the low-density outskirts, keeping the resulting $\pi^0$-decay gamma-ray
emission well below current detectability.

\section{Summary and Discussion}
\label{sec:s7}

We have investigated whether reconnection-driven injection can provide a
suprathermal seed population during cosmological magnetogenesis prior to
nonlinear structure formation. Motivated by the fact that reconnection
acceleration can be faster than instability-mediated stochastic acceleration,
we introduced a phenomenological source term $Q_{\rm rec}(p,z)$ into the
momentum-space Fokker--Planck equation. The source normalization was tied to a
fraction $\eta_{\rm rec}$ of the magnetic-energy growth rate, thereby linking
particle injection directly to the magnetic-energy budget of
pressure-anisotropy-driven magnetogenesis.

The timescale analysis shows that reconnection-driven injection can overcome
Coulomb cooling during the later stages of magnetic-field amplification. For
the fiducial effective efficiency $\eta_{\rm rec}=10^{-8}$, the condition
$t_{\rm rec,inj}=t_{\rm loss}$ is reached at $z_{\rm crit}\sim 5$--6 for the
representative low-energy particles considered here. Equivalently, the
critical efficiency $\eta_{\rm rec,crit}(E,z)$ decreases rapidly toward lower
redshift as the magnetic field grows, indicating that reconnection injection
becomes most viable only after significant magnetization has developed.

Solving the Fokker--Planck equation confirms this picture. In the absence of
the reconnection source, the high-momentum tail remains strongly suppressed.
When $Q_{\rm rec}$ is included, a suprathermal component develops above the
reconnection injection scale, and its normalization increases with
$\eta_{\rm rec}$. Thus, reconnection-driven injection can populate the
suprathermal regime and provide seed particles that are not produced by
stochastic acceleration alone.
However, the resulting seed population remains energetically small. Using the
suprathermal energy fraction defined above, we find
$u_{\rm nth}/u_{\rm th}\sim 10^{-7}$ for the fiducial case
$\eta_{\rm rec}=10^{-8}$. This is far below the typical fraction of
kinetic-energy flux converted into cosmic rays at structure-formation shocks,
which is expected to be of order $\sim 10^{-3}$--$10^{-1}$ depending on the
shock Mach number and injection physics. Reconnection-driven injection can
therefore produce a visible seed tail, but it is unlikely to make a
dynamically important contribution to the cosmic-ray energy budget before
nonlinear structure formation.

This limitation is a direct consequence of the extremely high-$\beta$ nature
of the pre-structure plasma. Since the reconnection source is powered by the
magnetic-energy growth rate, $\dot u_{\rm rec}\propto
\eta_{\rm rec}|dU_B/dt|$, the available energy reservoir is small when
$U_B=P_{\rm th}/\beta\ll P_{\rm th}$. Thus, even if reconnection provides a
fast injection channel, the total energy transferred to suprathermal particles
is strongly limited by the small magnetic-energy fraction.
The resulting tail is therefore not dynamically important by itself, but it
can still be kinetically relevant as a seed reservoir for later
structure-formation shocks. A test-particle shock reprocessing estimate shows
that a strong shock with $q=4$ can enhance the suprathermal proton population
above the adopted shock injection momentum by roughly an order of magnitude
relative to the stochastic-only case. Even so, the associated hadronic
gamma-ray emission from low-density cluster outskirts is expected to remain
far below current detectability. The main conclusion is that reconnection
during cosmological magnetogenesis primarily affects the seed supply for later
shock acceleration, rather than directly producing the dominant cosmic-ray
energy budget or an observable high-energy emission component.

\section*{Acknowledgements}
The numerical calculations presented in this work were performed using
computing resources at the Korea Astronomy and Space Science Institute
(KASI).

\bibliographystyle{elsarticle-harv} 
\bibliography{references}






\end{document}